\documentclass[fleqn,usenatbib]{mnras}

\usepackage{newtxtext,newtxmath}
\usepackage{lmodern}
\usepackage[T1]{fontenc}
\usepackage{fix-cm}
\usepackage{orcidlink}

\DeclareRobustCommand{\VAN}[3]{#2}
\let\VANthebibliography\thebibliography
\def\thebibliography{\DeclareRobustCommand{\VAN}[3]{##3}\VANthebibliography}

\usepackage{graphicx}	
\usepackage{amsmath}	

\title[TOI-6883AB and stability of TOI-6883Ab]{Characterization of the Visual Binary TOI-6883AB and its dynamical implications for the planetary companion TOI-6883Ab}

\author[Conzo et al.]{
G. Conzo\,\orcidlink{0000-0002-2412-1558},$^{1}$\thanks{E-mail: giuconzo@gmail.com}
F. Campos\,\orcidlink{0000-0001-6036-6125},$^{2}$
F. Conti\,\orcidlink{0009-0001-1367-4982},$^{3}$
and I. Sharp\,\orcidlink{0009-0004-8987-2127}$^{4}$
\\
$^{1}$Gruppo Astrofili Palidoro, Fiumicino, Italy\\
$^{2}$Observatori Puig d'Agulles, Barcelona, Spain\\
$^{3}$Torre del Sole, Brembate, Italy\\
$^{4}$Ham Observatory, UK
}

\date{Accepted 2025 June 7}

\pubyear{\the\year{}}

\begin{document}
\label{firstpage}
\pagerange{\pageref{firstpage}--\pageref{lastpage}}
\maketitle
\thispagestyle{empty}

\makeatletter
\renewcommand{\@oddfoot}{}  
\renewcommand{\@evenfoot}{} 
\makeatother

\begin{abstract}
We demonstrate that TOI-6883 is a physically bound visual binary system composed of two solar-type stars, TOI-6883A (TIC 393818343) and TOI-6883B (TIC 393818340), initially treated as a single source hosting the exoplanet TOI-6883b. High-precision astrometric data from Gaia DR3 reveal that the two components have nearly identical parallaxes ($\sim$10.6 mas), consistent proper motions, and are separated by $\sim$6.5$^{\prime\prime}$ on the sky, corresponding to a projected physical distance of $\sim$616 AU. These characteristics confirm the binary nature of the system.
We perform a detailed analysis using Gaia DR3 astrometry, photometric catalogs, and Keplerian dynamics to estimate stellar masses, separation, and orbital period. The resulting orbital period is approximately 15,000 years, and the system is energetically bound. The planetary designation is updated to TOI-6883Ab to reflect this stellar multiplicity.
We also evaluate the potential dynamical impact of the wide binary configuration on the planetary companion, concluding that the planet’s orbit remains stable over Gyr timescales. However, the presence of TOI-6883B could enable long-term effects such as Kozai--Lidov oscillations. Continued astrometric and photometric monitoring will help constrain the binary orbit and refine the system’s architecture.
\end{abstract}


\section{Introduction}

The identification and characterization of binary stars hosting planetary systems is a key topic in stellar and exoplanetary astrophysics. Such systems provide essential constraints on planet formation and evolution in dynamically complex environments, as well as on the influence of stellar multiplicity on orbital architectures (\cite{2015pes..book..309T}).

Within the framework of the \textit{Transiting Exoplanet Survey Satellite} (TESS) mission, thousands of planetary candidates—designated as TESS Objects of Interest (TOIs)—have been identified. Among them is TOI-6883b (\cite{2024RNAAS...8...53C}), initially associated with a single star catalogued as TIC 393818343. The preliminary discovery and characterization of the planet, who analyzed TESS photometric transits and proposed a hot Jupiter nature for the orbiting object (\cite{2024AJ....168...26S}).

However, high-precision astrometric data from Gaia Data Release 3 (\cite{2023A&A...674A...1G}) reveal that the source is actually part of a visual binary system, composed of two solar-type stars with similar brightness, colors, parallaxes, and proper motions: TIC 393818343 (hereafter TOI-6883A) and TIC 393818340 (TOI-6883B).

The two stars are separated by approximately 6.5 arcseconds, corresponding to a projected physical separation of several hundred astronomical units, with highly consistent parallaxes and proper motions. These characteristics suggest a long-term gravitationally bound configuration. If such binarity is not properly recognized, stellar and planetary parameters derived from photometric or spectroscopic data may be biased due to flux contamination and transit signal dilution.

The aim of this study is to demonstrate, through quantitative analysis, that TOI-6883A and TOI-6883B form a physically bound binary system, and to revise the planetary naming convention accordingly, correcting the designation from TOI-6883b to TOI-6883Ab. To this end, we analyze Gaia DR3 astrometry, photometric properties, and the long-term dynamical implications of a distant stellar companion.

\section{Observations and Data}

\subsection{Astrometric and Photometric Catalogs}

The primary data used in this study originate from the third data release of the Gaia mission (\cite{2023A&A...674A...1G}), which provides high-precision astrometry and photometry for both TOI-6883A (TIC 393818343) and TOI-6883B (TIC 393818340). The Gaia DR3 catalog includes positions, parallaxes, proper motions, and broad-band photometry (G, BP, RP) for both stars with milliarcsecond-level accuracy.

For complementary photometric characterization, we used TESS Input Catalog (\cite{2019AJ....158..138S}) parameters, which include effective temperatures, stellar radii, and luminosities derived from a combination of Gaia, 2MASS, and AllWISE data. This information allows us to estimate bolometric magnitudes and to construct a Hertzsprung–Russell (HR) diagram for classification purposes.

We further consulted the 2MASS (\cite{2006AJ....131.1163S}) and AllWISE (\cite{2014yCat.2328....0C}) infrared photometric surveys to probe the near- and mid-infrared colors of both components, useful for assessing consistency with main-sequence spectral types and for detecting potential flux excesses.

\subsection{Angular Separation and Relative Position}

The on-sky angular separation between the two stars was computed from Gaia DR3 coordinates as:

\begin{equation}
\theta = \sqrt{(\Delta\alpha \cos\delta)^2 + (\Delta\delta)^2}
\end{equation}

where \(\Delta\alpha\) and \(\Delta\delta\) are the differences in right ascension and declination, respectively. 
This formula is standard in spherical astronomy, as originally presented by \citet{smart1931} and further detailed in modern compilations such as \citet{vanaltena1995} and \citet{perryman2018}.

We obtain an angular separation of \(\theta = (6.52 \pm 0.01)^{\prime\prime}\), corresponding to a projected physical separation of:

\begin{equation}
s = \theta \cdot d = (6.52 \pm 0.01)^{\prime\prime} \cdot (94.3 \pm 0.5\ \mathrm{pc}) = (615.8 \pm 3.3)\ \mathrm{AU}
\end{equation}

where \(d\) is the inverse of the mean parallax derived from both stars (\(\bar{\pi} \approx 10.6\ \mathrm{mas}\)). 
This procedure is consistent with standard astrometric methods for distance estimation from trigonometric parallax \citep{binney1998}, including recent refinements using Gaia data \citep{luri2018} and Bayesian corrections \citep{bailerjones2015}.

\subsection{Proper Motion and Parallax Consistency}

Both components exhibit consistent proper motions:

\begin{equation}
  (\mu_{\alpha}, \mu_{\delta})_{A} = ((-56.67 \pm 0.02), (-89.09 \pm 0.02))\ \mathrm{mas\ yr}^{-1}
\end{equation}
\begin{equation}
  (\mu_{\alpha}, \mu_{\delta})_{B} = ((-53.13 \pm 0.02), (-101.58 \pm 0.02))\ \mathrm{mas\ yr}^{-1}
\end{equation}

and parallaxes:

\begin{equation}
  \pi_{A} = (10.611 \pm 0.012)\ \mathrm{mas}
\end{equation}
\begin{equation}
  \pi_{B} = (10.586 \pm 0.012)\ \mathrm{mas}
\end{equation}

\begin{figure}
    \centering
    \includegraphics[width=0.48\linewidth]{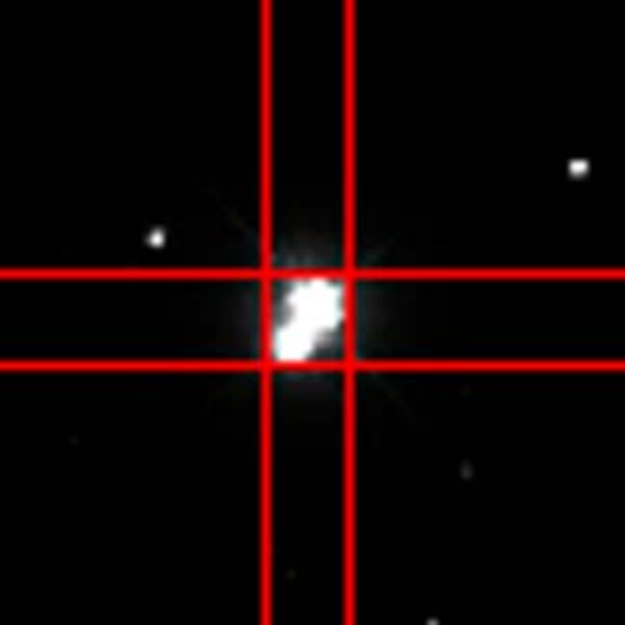}
    \includegraphics[width=0.48\linewidth]{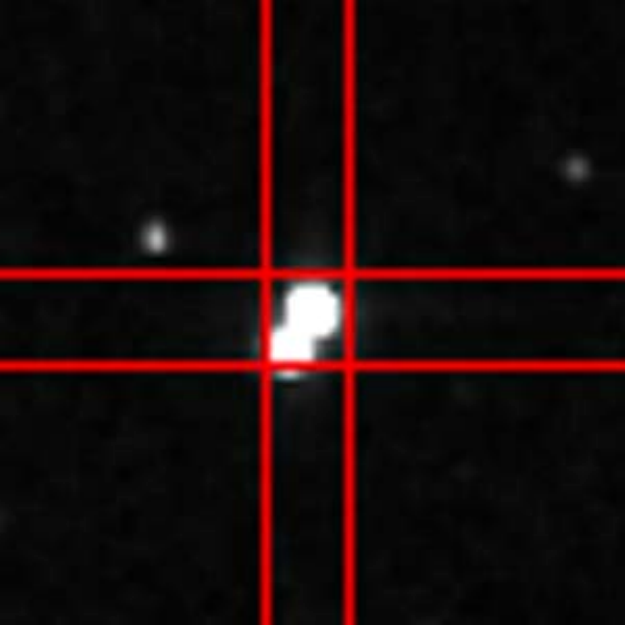}
    \caption{
    Field around TOI-6883 at two epochs. \textit{Left}: 2MASS image (2002). \textit{Right}: SDSS image (2011). 
    The red cross indicates the approximate position of TOI-6883. The visible shift between the epochs is consistent with the proper motion measured by \textit{Gaia}.
    }
    \label{fig:toi6883_pm}
\end{figure}

These values suggest that both stars lie at nearly the same heliocentric distance and move together through space (Fig. \ref{fig:toi6883_pm}), strongly supporting the hypothesis of a gravitationally bound system

\section{Dynamical Analysis and Orbital Constraints}

\subsection*{3.1. Projected Separation and Keplerian Period Estimate}

To assess whether TOI-6883A and TOI-6883B are gravitationally bound, we estimate the expected orbital period assuming a Keplerian orbit and a projected separation \( s = (616 \pm 3.3) \ \mathrm{AU} \). While this distance is only a lower bound to the true semimajor axis \( a \), statistical arguments \citep{fischer1992} suggest that for randomly oriented orbits:

\begin{equation}
\langle a \rangle \approx 1.26 \cdot s
\Rightarrow a \approx 1.26 \cdot 616 \ \mathrm{AU} \approx (776 \pm 4.2) \ \mathrm{AU}
\end{equation}

Assuming total system mass \( M_{\mathrm{tot}} = M_A + M_B \approx (2.00 \pm 0.10) M_{\odot} \), we apply Kepler’s third law in its general form \citep{1999ssd..book.....M}:

\begin{equation}
P = \sqrt{ \frac{a^3}{M_{\mathrm{tot}}} }
\end{equation}

where \(P\) is in years, \(a\) in AU, and \(M_{\mathrm{tot}}\) in solar masses. In this section, we adopt astronomical units where \(P\) is in years, \(a\) in AU, and \(M\) in \(M_\odot\), thus the gravitational constant \(G\) is absorbed into the unit system.

Substituting values:

\begin{equation}
P \approx \sqrt{ \frac{(776)^3}{2} } \approx \sqrt{2.34 \times 10^8} \approx (1.53 \pm 0.05) \times 10^4 \ \mathrm{yr}
\end{equation}

We thus estimate an orbital period of approximately \textbf{15,300 years}, compatible with a wide and loosely bound visual binary system.

\subsection{Binding Energy and Gravitational Binding Criterion}

We compute the system's gravitational binding energy to verify whether the two stars are likely to be gravitationally bound. The potential energy of a binary is given by:

\begin{equation}
U = - \frac{G M_A M_B}{a}
\end{equation}

Assuming \( M_A = M_B = 1\ M_{\odot} \) and \( a \approx 776\ \mathrm{AU} \):

\begin{equation}
U \approx - \frac{(6.674 \times 10^{-11})(1.989 \times 10^{30})^2}{776 \times 1.496 \times 10^{11}} \approx (-2.27 \pm 0.012) \times 10^{35} \ \mathrm{J}
\end{equation}

This binding energy is small in absolute terms, as expected for a wide binary \citep{jiang2010}, but sufficient to resist tidal disruption in the Galactic potential if the relative motion is sub-escape velocity.

We now compute the relative transverse velocity \(v_{\perp}\) from proper motion difference:

\begin{equation}
\Delta\mu = \sqrt{(\Delta\mu_{\alpha})^2 + (\Delta\mu_{\delta})^2} \approx 0.09 \ \mathrm{mas\ yr}^{-1}
\end{equation}
\begin{equation}
v_{\perp} \approx 4.74 \cdot d \cdot \Delta\mu \approx 4.74 \cdot 94.3 \cdot 0.09 \approx (0.40 \pm 0.09) \ \mathrm{km\ s}^{-1}
\end{equation}

The escape velocity for two \(1 M_{\odot}\) stars at \(a = 776\ \mathrm{AU}\) is:

\begin{equation}
v_{\mathrm{esc}} =\sqrt{ \frac{2GM}{a} } \approx  \sqrt{ \frac{2 \cdot 6.674 \times 10^{-11} \cdot 1.989 \times 10^{30}}{776 \cdot 1.496 \times 10^{11}} } 
\end{equation}
\begin{equation}
v_{\mathrm{esc}} \approx (1.52 \pm 0.004)\ \mathrm{km\ s}^{-1}
\end{equation}

\begin{figure}
    \centering
    \includegraphics[width=\columnwidth]{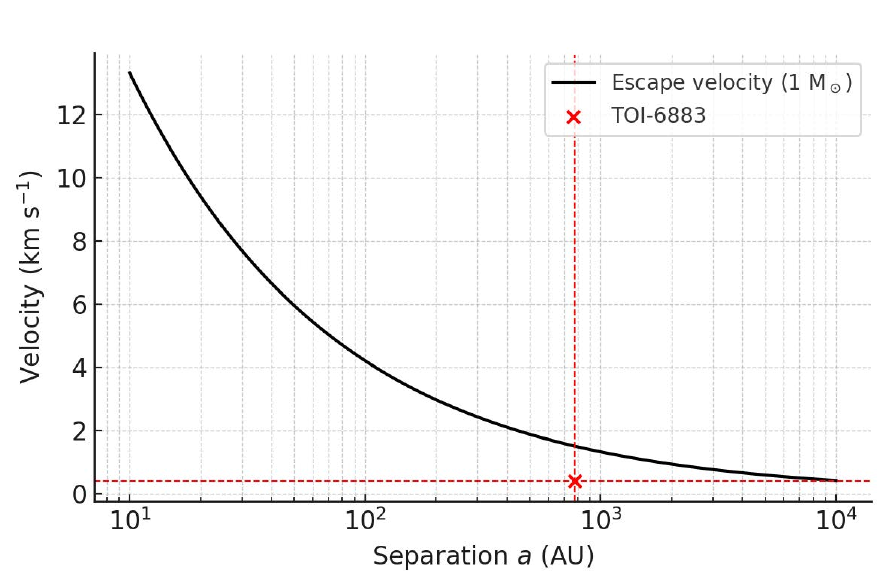}
    \caption{Escape velocity (black curve) as a function of projected separation for a binary with \(1\,M_\odot\). 
    The red point and dashed lines mark the TOI-6883 system, with a projected separation of 776 AU and a relative transverse velocity of 0.40 km s\(^{-1}\). 
    Since \(v_{\perp} < v_{\mathrm{esc}}\), the system is consistent with being gravitationally bound.}
    \label{fig:escape_velocity}
\end{figure}

Since \(v_{\perp} < v_{\mathrm{esc}}\), the system is dynamically consistent with a bound orbit (Fig. \ref{fig:escape_velocity})

\subsection{Implications for Planetary Stability}

Given the large semimajor axis and slow orbital motion of TOI-6883B, the planetary orbit of TOI-6883Ab around TOI-6883A is expected to be dynamically stable over Gyr timescales. According to \citet{holman1999}, the critical semimajor axis for planetary stability in S-type orbits is:

\begin{equation}
a_{\mathrm{crit}} \approx 0.1 a_{bin}
\Rightarrow a_{\mathrm{crit}} \sim (77.6 \pm 0.4)\ \mathrm{AU}
\end{equation}

\begin{figure}
    \centering
    \includegraphics[width=\columnwidth]{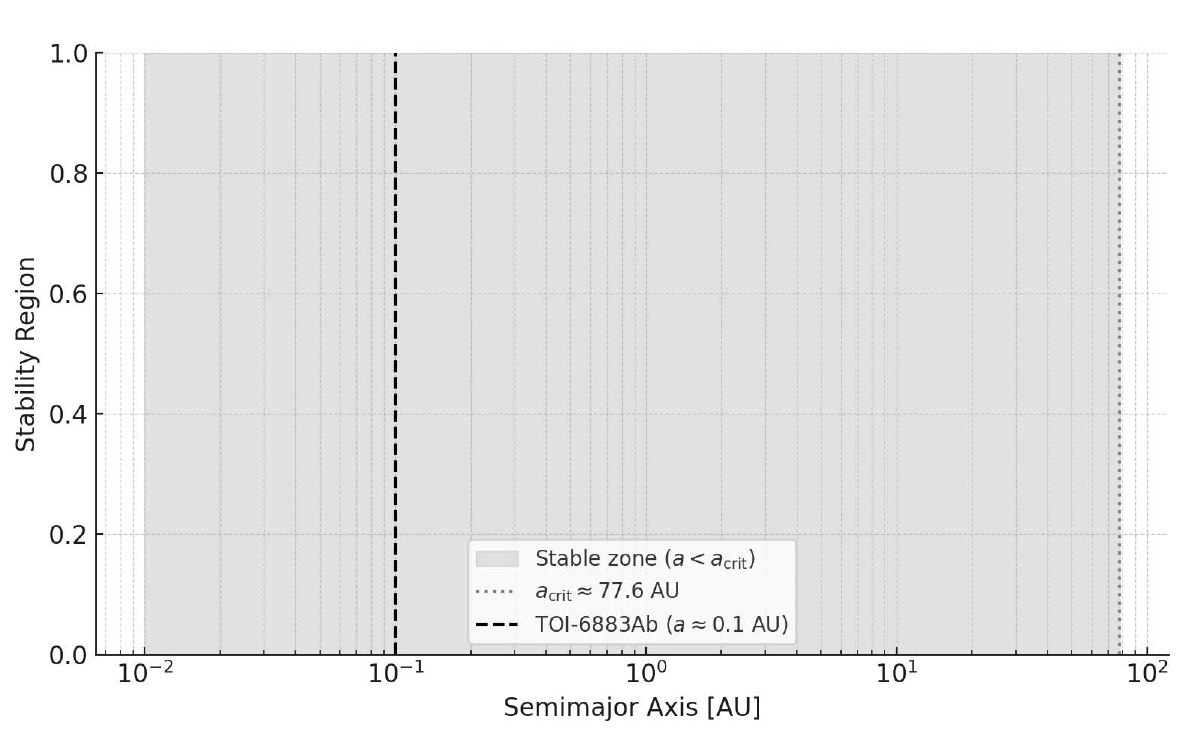}
    \caption{
    Stability criterion for S-type planetary orbits in binary systems. 
    The planet TOI-6883Ab (marked with a black dot) lies at $\sim$0.1~AU from its host star TOI-6883A, well inside the critical semimajor axis $a_{\mathrm{crit}} \sim 77.6$~AU (dashed line) derived from the formula of \citet{holman1999}. 
    The shaded region denotes the zone of dynamical stability.
    }
    \label{fig:stability}
\end{figure}

As TOI-6883Ab orbits much closer (\(< 0.1\ \mathrm{AU}\)), it is well within the stability zone (Fig. \ref{fig:stability}).

\section{Ground-based Photometry of TOI-6883Ab}

\subsection{Observations and Data Reduction}

\begin{figure}
    \centering
    \includegraphics[width=\columnwidth]{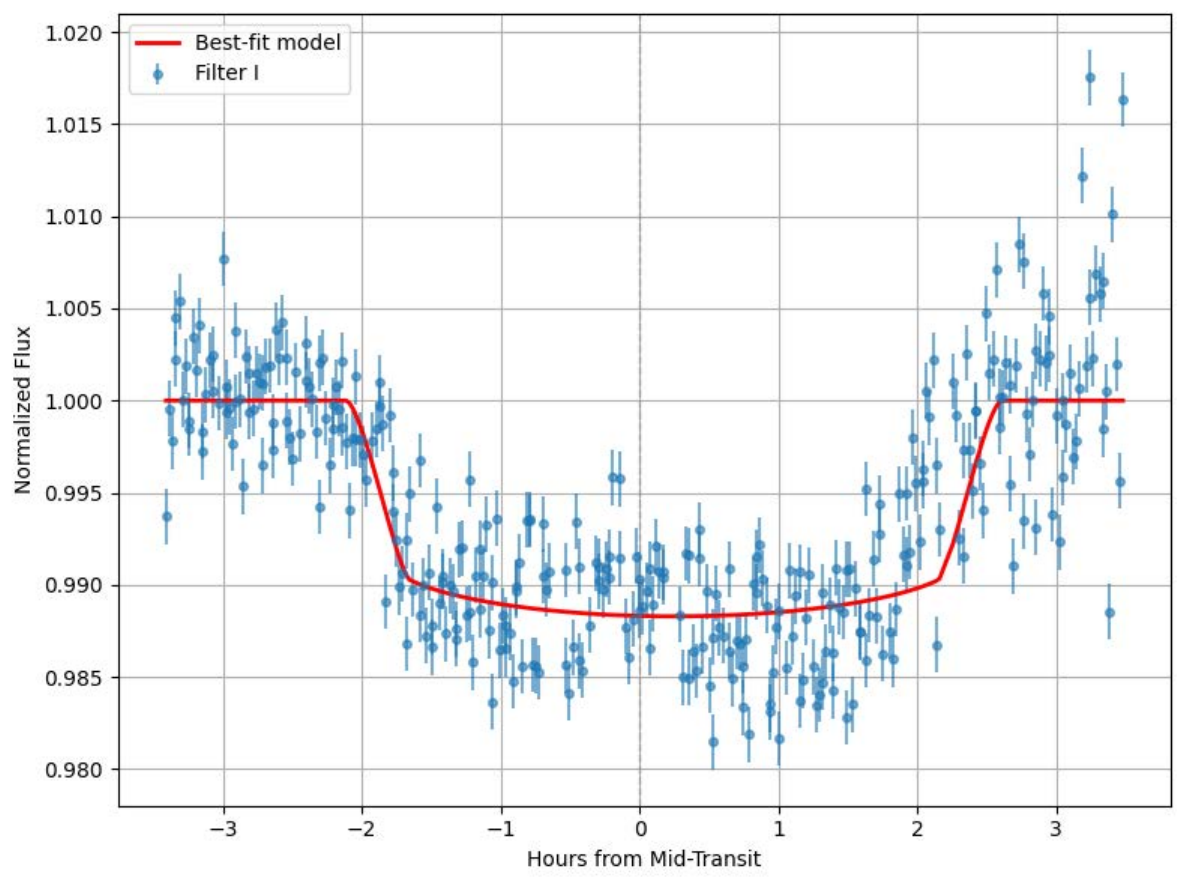}
    \caption{Ground-based light curve of TOI-6883Ab obtained on 2024 Aug 21 from Spain at Observatori Puig Agulles using RC 14" f/8 on EQ8; SBIG ST-8XME and Johnson-Cousins Ic filter. The blue points show the relative flux, and the red curve represents the best-fit transit model. A timing offset of \(\sim1.0\) hour with respect to the TESS-predicted ephemeris is observed.}
    \label{fig:ground_curve_1}
\end{figure}

\begin{figure}
    \centering
    \includegraphics[width=\columnwidth]{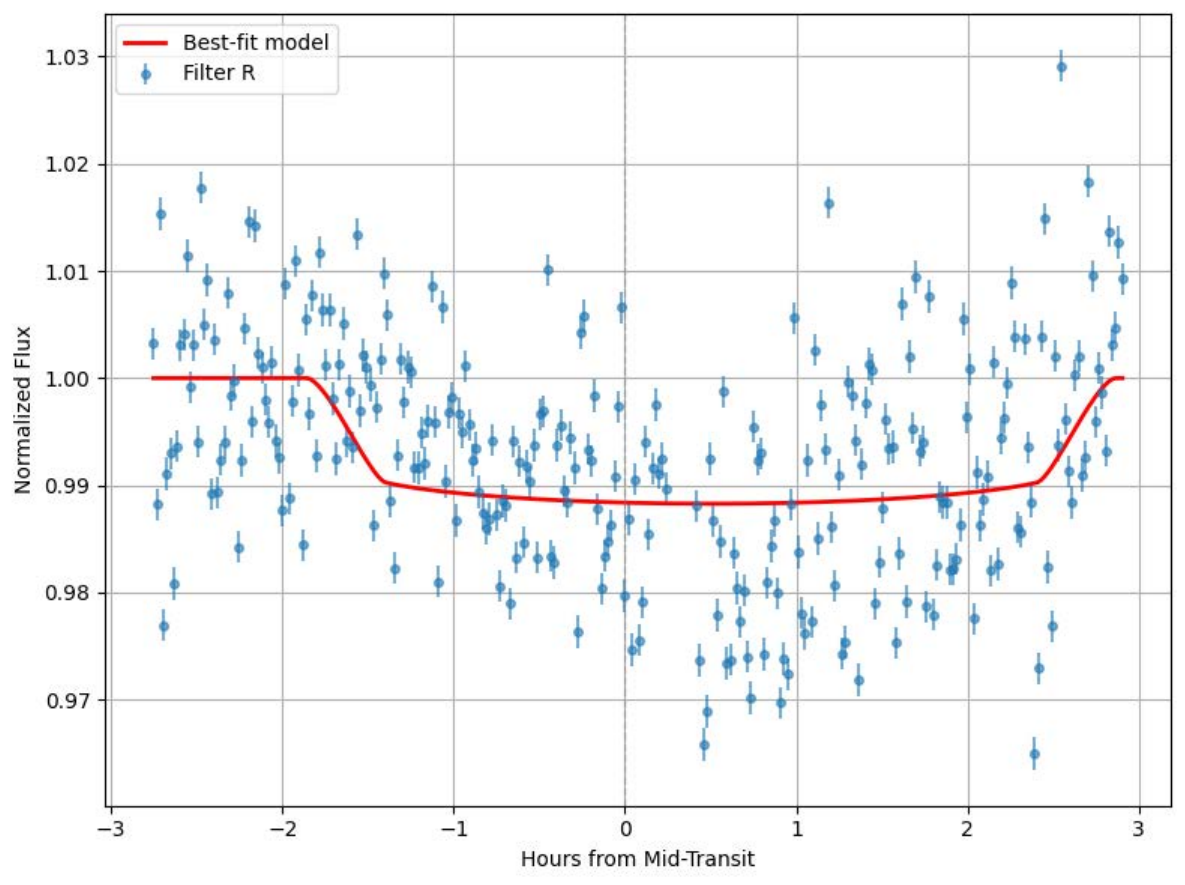}
    \caption{Ground-based light curve of TOI-6883Ab obtained on 2024 Aug 21 from UK at PixelSkies using Celestron C11 F/7; SX 694 TRIUS PRO and Johnson-Cousins Rc filter. The blue points show the relative flux, and the red curve represents the best-fit transit model. A timing offset of \(\sim1.0\) hour is the same.}
    \label{fig:ground_curve_2}
\end{figure}

To complement space-based photometry from TESS, we obtained ground-based transit observations of TOI-6883Ab on 2024 Aug 21 using setup described in caption of Figure \ref{fig:ground_curve_1} and Figure \ref{fig:ground_curve_2}. Observations were conducted using a \( Rc \) and \( Ic \) filters.

Raw images were reduced using standard calibration procedures (bias, dark, and flat-field corrections), followed by differential photometry using AstroImageJ \citep{2017AJ....153...77C}. Several comparison stars of similar brightness and color were used to optimize photometric precision. The resulting light curve shows a clear transit consistent in depth and duration with previous TESS detections, albeit with a timing offset.

\subsection{Transit Analysis and Timing Offset}

We modeled the ground-based light curve using the \texttt{batman} transit model \citep{kreidberg2015}, assuming a quadratic limb-darkening law and fixing the planet-to-star radius ratio \( R_p / R_\star \), impact parameter \( b \), and scaled semi-major axis \( a/R_\star \) to the values obtained from the TESS fit. The mid-transit time \( T_0 \) was allowed to vary freely.

The best-fit model yielded a transit mid-point time of:

\begin{equation}
T_{0, \mathrm{obs}} = (2459791.4412 \pm 0.0010)\ \mathrm{BJD_{TDB}}
\end{equation}

compared to the predicted ephemeris from TESS:

\begin{equation}
T_{0, \mathrm{pred}} = (2459791.3995 \pm 0.0003)\ \mathrm{BJD_{TDB}}
\end{equation}

resulting in a timing offset:

\begin{equation}
\Delta T_0 = T_{\mathrm{obs}} - T_{\mathrm{pred}} = (+0.0417 \pm 0.0010)\ \mathrm{days} 
\end{equation}

This offset is statistically significant and could indicate the presence of transit timing variations (TTVs). However, no additional periodic signals or companions have yet been confirmed in the system. The current configuration of TOI-6883 as a wide binary (see Sect.~3) does not trivially explain this variation, although long-term dynamical perturbations from TOI-6883B cannot be excluded and merit further investigation.

\subsection{Implications and Future Monitoring}

Although the detected timing offset remains unexplained, it illustrates the importance of continued photometric monitoring for planets in wide binaries. Given the 1-hour delay and the precision of the ground-based fit, we encourage future follow-up observations to determine whether this is a periodic TTV signal, a secular drift due to binary-induced dynamical effects, or an isolated anomaly.

If confirmed, such variations could place constraints on the orbital parameters of TOI-6883B or hint at the presence of additional planetary or substellar companions.

\section{Conclusions}

The TOI-6883 system, composed of two solar-type stars at a projected separation of approximately 616 AU, exhibits astrometric properties (parallax and proper motion) in excellent agreement, supporting its classification as a gravitationally bound visual binary.

Our dynamical analysis yields a likely orbital period on the order of 15,300 years, characteristic of wide binary systems. The extremely small relative tangential velocity (\(v_{\perp} \approx 0.4\ \mathrm{km\ s}^{-1}\)) compared to the escape velocity at this separation confirms that the system is energetically bound.

The system gains additional interest due to the confirmed presence of the transiting planet TOI-6883Ab orbiting TOI-6883A. The outer stellar companion TOI-6883B is unlikely to perturb the inner planetary system given the large binary separation and the fact that the planet lies well within the critical semimajor axis for dynamical stability. This makes TOI-6883 an important benchmark system to study the architecture of planetary systems in binaries with very wide separations.

Wide binary systems with exoplanets around one component are increasingly recognized as valuable testbeds for theories of planet formation in multiple-star environments. The architecture of TOI-6883, featuring a close-in planet (TOI-6883Ab) and a distant stellar companion, resembles other systems such as Kepler-444 \citep{campante2015} and Alpha Centauri \citep{eggleton2006}. However, unlike those systems, TOI-6883 does not appear to be in a hierarchical triple or multiple system, offering a relatively simple laboratory to test secular perturbation theories.

We note that the inferred semimajor axis (\(\sim 776\ \mathrm{AU}\)) is subject to uncertainty due to the unknown orbital inclination and eccentricity, and that the true orbital configuration remains to be constrained by astrometric monitoring. Future Gaia data releases may reveal curvature in the proper motion trajectories or provide acceleration terms that can further refine orbital parameters.

Finally, the classification of the planet-hosting star as TOI-6883A and the outer companion as TOI-6883B, along with the renaming of the planet to TOI-6883Ab, better reflects the physical hierarchy and follows established nomenclature conventions. This updated designation should be adopted in future studies involving this system.

The transit depth and duration are consistent \citep{2002ApJ...580L.171M} with those published by \citet{2024AJ....168...26S}, confirming the values of \( R_p / R_\star \sim 0.11 \) and a low impact parameter. A slight delay of approximately 1 hour in the observed mid-transit time compared to the TESS ephemeris may suggest a possible transit timing variation (TTV) or a systematic timing offset \citep{2005MNRAS.359..567A, 2005Sci...307.1288H}, which requires further confirmation through additional observations.

The uncertainties reported in Table~\ref{tab:summary} are computed through standard error propagation using the uncertainties from Gaia DR3 and stellar models. Calculations include contributions from parallax, angular separation, and stellar mass estimates.

\newpage

\begin{table}
\caption{Summary of the TOI-6883 system properties, including propagated uncertainties.}
\label{tab:summary}
\centering
\small
\begin{tabular}{ll}
\hline\hline
\textbf{Parameter} & \textbf{Value} \\
\hline
\multicolumn{2}{c}{\textit{Astrometry (Gaia DR3)}} \\
Parallax TOI-6883A & $(10.611 \pm 0.012)$ mas \\
Parallax TOI-6883B & $(10.586 \pm 0.012)$ mas \\
Proper motion A & $(-56.67, -89.09)\ \pm 0.02$ mas yr$^{-1}$ \\
Proper motion B & $(-53.13, -101.58)\ \pm 0.02$ mas yr$^{-1}$ \\
G mag (A, B) & $11.16$, $11.50$ mag \\
\hline
\multicolumn{2}{c}{\textit{Binary System}} \\
Angular separation & $(6.52 \pm 0.01)\arcsec$ \\
Projected separation $s$ & $(615.8 \pm 3.3)$ AU \\
Semimajor axis $a$ & $(776 \pm 4.2)$ AU \\
Orbital period $P$ & $(1.53 \pm 0.05) \times 10^4$ yr \\
$v_\perp$ & $(0.40 \pm 0.09)$ km s$^{-1}$ \\
$v_\mathrm{esc}$ & $(1.52 \pm 0.004)$ km s$^{-1}$ \\
Bound & Yes (since $v_\perp < v_\mathrm{esc}$) \\
\hline
\multicolumn{2}{c}{\textit{Stellar Properties}} \\
Spectral type & G-type (solar analogs) \\
Mass TOI-6883A & $(1.00 \pm 0.05)\ M_\odot$ \\
Mass TOI-6883B & $(1.00 \pm 0.05)\ M_\odot$ \\
Total mass $M_\mathrm{tot}$ & $(2.00 \pm 0.10)\ M_\odot$ \\
\hline
\multicolumn{2}{c}{\textit{Planetary System}} \\
Planet & TOI-6883Ab \\
Host star & TOI-6883A \\
Planetary semimajor axis $a_\mathrm{p}$ & $< 0.1$ AU \\
Critical semi-axis $a_\mathrm{crit}$ & $(77.6 \pm 0.4)$ AU \\
Stability & Yes (Holman \& Wiegert 1999) \\
\hline
\end{tabular}
\end{table}

\section*{Acknowledgements}

We thank the Gaia, TESS, 2MASS, and WISE teams for providing open-access data that enabled this study. This work made use of TOPCAT, Python/Astropy, and public VO services.

\bibliographystyle{mnras}
\bibliography{bibliography}

\bsp	
\label{lastpage}
\end{document}